\documentclass[aps,preprint]{revtex4}%
\usepackage{amsfonts}%
\usepackage{amsmath}%
\usepackage{amssymb}%
\usepackage{graphicx}

\begin{document}
\preprint{ }
\title[measuring energy dissipation rate in anisotropic turbulence]{Methods for measuring energy dissipation rate in anisotropic turbulence}
\author{Reginald J. Hill}
\affiliation{Cooperative Institute for Research in Environmental Sciences, University of
Colorado/NOAA Earth System Research Laboratory, Physical Sciences Division,
325 Broadway, Boulder, CO 80305-3337}

\begin{abstract}
Energy dissipation rate, $\varepsilon$, is an important parameter for nearly
every experiment on turbulent flow. \ Mathematically precise relationships
between $\varepsilon$ and other measurable statistics for the case of
anisotropic turbulence are useful to experimentalists. \ Such relationships
are obtained for which the measurable statistics are the 3rd-order and
2nd-order velocity structure functions as well as the acceleration-velocity
structure function. \ The relationships are derived using the Navier-Stokes
equation without approximation. \ Approximate versions are obtained on the
basis of local stationarity and local homogeneity. \ The latter are valid for
arbitrary Reynolds numbers for the case of stationary, homogeneous turbulence.
\ Precise use of the mathematics requires care noted in the Discussion section.

\end{abstract}
\maketitle

\section{Introduction}

\qquad Recently, many experiments have produced turbulence by novel means
and/or measured energy dissipation rate $\varepsilon$\ by novel means. \ Luthi
\textit{et al.} (2005), use a magnetically driven flow that is nearly
isotropic, and obtain $\varepsilon$ from velocity derivatives including
$2\nu\left\langle s_{ij}s_{ij}\right\rangle $, where $s_{ij}$ is the rate of
strain, $\nu$ is kinematic viscosity, and repeated indices denote summation
over coordinate directions. \ Mann \textit{et al.} (1999) and Ott \& Mann
(2005) use an oscillating grid, particle tracking, and obtain $\varepsilon$
from $\left\langle \left(  a_{i}-a_{i}^{\prime}\right)  \left(  u_{i}%
-u_{i}^{\prime}\right)  \right\rangle $, where $a_{i}$ and $u_{i}$ are
acceleration and velocity. \ Similarly, Ott \& Mann (2000) compare those
methods with $\varepsilon$ determined from the 3rd- and 2nd-order structure
function. \ Voth \textit{et al.} (2002) and La Porta \textit{et al.} (2001)
generate turbulence between counter-rotating blades in a cylindrical
enclosure; using particle tracking, they obtain $\varepsilon$ from 2nd-order
structure functions. \ Ouellette \textit{et al.} (2006) create the same
turbulence and include evaluation of $\varepsilon$ from $\left\langle \left(
a_{i}-a_{i}^{\prime}\right)  \left(  u_{i}-u_{i}^{\prime}\right)
\right\rangle $. \ Berg \textit{et al.} (2006) use multiple propellers to
create turbulence, use particle tracking, and obtain $\varepsilon$ from
2nd-order velocity structure functions. \ Tsinober \textit{et al.} (1992),
Kolmyansky \textit{et al.} (2001), and Gulitski \textit{et al.} (2007) measure
in the atmospheric surface layer using multi-wire probes. \ Without use of
Taylor's frozen flow hypothesis they obtain $\varepsilon$ from velocity
derivatives including $2\nu\left\langle s_{ij}s_{ij}\right\rangle $.

\qquad A common means of obtaining $\varepsilon$\ is from the inertial range
of the energy spectrum or that of the 2nd-order velocity structure function.
\ That relationship is based on dimensional analysis under the assumption of
local isotropy, and based on empirical validation, and on empirical evaluation
of the Kolmogorov constant. \ For anisotropic turbulence, that empirical basis
must be reevaluated for each anisotropic flow. \ Those relationships have not
been derived from the Navier-Stokes equation, and are therefore not considered here.

\qquad The usefulness of asymptotic relationships for an experimenter's
purpose of quantifying $\varepsilon$\ using measurements of other statistics
is limited. \ For every different type of flow and every Reynolds number, the
experimenter must demonstrate to what accuracy the asymptotic limit permits
determination of $\varepsilon$. \ Duchon \& Robert (2000) and Eyink (2003) use
time and space averages of arbitrary extent and $\nu=0$ such that the Reynolds
number is infinite and use an orientation average to remove the effect of
anisotropy. \ Both the time and space averages must be of nonvanishing extent
(Eyink (2003) and \S 6.2 in Hill (2006)). \ The requirement $\nu=0$ is not
applicable to experiments. \ Eyink (2003) considers the possibility of
experimental tests of their results, concluding that \textquotedblleft a slow
approach to asymptopia makes a direct test of the local results, especially a
verification of the numerical prefactor, rather more
difficult.\textquotedblright\ \ A likely cause of inaccuracy in experiments of
the results of Duchon \& Robert (2000) and Eyink (2003) is an effect of random
sweeping described in \S 6.1 in Hill (2006).

\qquad The time and space averages used by Nie and Tanveer (1999) are not
applicable to experiments. \ Nie \& Tanveer (1999) base their relationship
between $\varepsilon$ and the 3rd-order velocity structure function on time
averaging over an infinite duration to remove effects of nonstationarity, on
space averaging over either all of space without boundary conditions or space
averaging over an entire spatial period of a spatially periodic flow to remove
effects of inhomogeneity, and on a sufficiently large Reynolds number. \ They
use an orientation averaging to remove the effect of anisotropy.

\qquad Danaila \textit{et al}. (2002, 2004, and references therein) present
structure function equations in which terms that describe particular effects
of large-scale inhomogeneity are retained. \ They use data to demonstrate that
those retained large-scale terms are the dominant inhomogeneous terms for
several types of flows. \ Their equations are approximate because the
large-scale inhomogeneity terms are approximated, and the other terms are
approximated using local isotropy.

\qquad In contrast, the method in this paper avoids approximation.\ \ On the
basis of algebra, calculus, incompressibility, the Navier-Stokes equation, and
use of no approximations whatsoever, the structure function equation that
contains energy dissipation rate is given in \S 4. \ The new, quantifiable
definitions of local homogeneity and local stationarity are illustrated in
\S 3.\ \ In \S 4, terms that are negligible for local homogeneity and local
stationarity are identified. \ For simplicity, but not for necessity, those
terms are not carried forward in (\ref{structure function})--(\ref{4a result}%
). \ Those terms are considered in \S 7. \ For simplicity, an ensemble average
is used herein because the slight complications from spatial and temporal
averages were thoroughly documented in Hill (2002a,b). \ With attention to
those complications, spatial and temporal averages may be substituted for the
ensemble average used here. \ An average conditioned on the value of some
hydrodynamic quantity cannot be used because of its unknown commutation with
respect to spatial and temporal derivatives; that topic must await future study.

\qquad Several methods of obtaining energy dissipation rate $\varepsilon
$\ without use of local isotropy are given here. \ These methods derive from
the Navier-Stokes equation with no restriction on the flow symmetry; they
differ only in the type of average employed.\ These methods are expected to be
of use to experimenters in interpretation of their data. \ Our purpose here is
to provide pragmatic means for measuring $\varepsilon$ using no assumptions
and with the greatest mathematical precision possible.\ \ Cautions to
experimenters on precise use of the relationships are in \S 7.

\section{Notation}

\qquad The velocities, accelerations, and energy dissipation rates at spatial
points $\mathbf{x}$\ and $\mathbf{x}^{\prime}$\ and times $t$\ and $t^{\prime
}$\ are denoted by%
\begin{equation}
u_{i}\equiv u_{i}(\mathbf{x},t),u_{i}^{\prime}\equiv u_{i}(\mathbf{x}^{\prime
},t^{\prime}),a_{i}\equiv a_{i}(\mathbf{x},t),a_{i}^{\prime}\equiv
a_{i}(\mathbf{x}^{\prime},t^{\prime}),\varepsilon\equiv\varepsilon
(\mathbf{x},t),\varepsilon^{\prime}\equiv\varepsilon(\mathbf{x}^{\prime
},t^{\prime}),
\end{equation}
etc.; $\mathbf{x}$, $t$, $\mathbf{x}^{\prime}$, $t^{\prime}$ are independent
variables. \ For particle tracking measurements, $u_{i}$\ could be the
velocity of one particle at position $\mathbf{x}$\ at time $t$,\ and
$u_{i}^{\prime}$\ could be the velocity of another particle at position
$\mathbf{x}^{\prime}$\ at time $t^{\prime}$, where $t=t^{\prime}$ or $t\neq
t^{\prime}$, but if $u_{i}$\ and $u_{i}^{\prime}$\ are the velocities of the
same particle at different $\mathbf{x}$ and $\mathbf{x}^{\prime}$, then
clearly $t\neq t^{\prime}$. \ A sequence of point pairs may come from particle
trajectories, but it is useful to consider them as pairs of coordinate
locations. \ Define a new set of independent variables:%

\begin{equation}
\mathbf{X}\equiv\left(  \mathbf{x}+\mathbf{x}^{\prime}\right)  /2\text{ \ and
\ }\mathbf{r}\equiv\mathbf{x}-\mathbf{x}\text{, \ and }r\equiv\left\vert
\mathbf{r}\right\vert \text{; \ \c{T}}\equiv\left(  t+t^{\prime}\right)
/2\text{ \ and \ \c{t}}\equiv t-t^{\prime}.\label{variable change}%
\end{equation}
The significance of variables $\mathbf{X}$ and \c{T} is that they are the
location and time of the measurement, respectively. \ Define the following:%
\begin{align}
d_{ij} &  \equiv\left(  u_{i}-u_{i}^{\prime}\right)  \left(  u_{j}%
-u_{j}^{\prime}\right) \label{dij}\\
d_{ijk} &  \equiv\left(  u_{i}-u_{i}^{\prime}\right)  \left(  u_{j}%
-u_{j}^{\prime}\right)  \left(  u_{k}-u_{k}^{\prime}\right) \\
A_{ij} &  \equiv\left(  a_{i}-a_{i}^{\prime}\right)  \left(  u_{j}%
-u_{j}^{\prime}\right)  +\left(  a_{j}-a_{j}^{\prime}\right)  \left(
u_{i}-u_{i}^{\prime}\right) \\
\digamma_{iik} &  \equiv\left(  u_{i}-u_{i}^{\prime}\right)  \left(
u_{i}-u_{i}^{\prime}\right)  \frac{u_{k}+u_{k}^{\prime}}{2}%
\end{align}
Below, we use numerical subscript $1$ to denote a component in the direction
of $\mathbf{r}$; and $2$ and $3$ to denote components transverse to
$\mathbf{r}$.

\section{What are local homogeneity and local stationarity?}

\qquad Local homogeneity and local stationarity only apply for very large
Reynolds numbers and sufficiently small $r$. \ Require that the definition of
local homogeneity produces the same results as homogeneity, and that the
approximation be quantifiable. \ The simplest case is the incompressibility
relationship on the 2nd-order structure function, namely the divergence
vanishes: $\partial_{r_{j}}\left\langle d_{ij}\right\rangle =0$, where we
denote an ensemble average by angle brackets. \ Familiar examples are obtained
by substitution of the isotropic formula for $\left\langle d_{ij}\right\rangle
$ into $\partial_{r_{j}}\left\langle d_{ij}\right\rangle =0$; that gives the
well-known incompressibility relationship that $r\partial_{r}\left\langle
d_{11}\right\rangle +2\left[  \left\langle d_{11}\right\rangle -\left\langle
d_{22}\right\rangle \right]  =0 $; subsequent substitution of the
inertial-range $2/3$ power-law formula for $\left\langle d_{ij}\right\rangle $
gives the familiar inertial-range result that $\left\langle d_{22}%
\right\rangle =\left(  4/3\right)  \left\langle d_{11}\right\rangle $, and
similarly substituting the viscous-range formula for $\left\langle
d_{ij}\right\rangle $ gives the familiar relationship $\left\langle \left(
\partial u_{2}/\partial x_{1}\right)  ^{2}\right\rangle =2\left\langle \left(
\partial u_{1}/\partial x_{1}\right)  ^{2}\right\rangle $. \ Use of algebra,
calculus, and incompressibility, i.e., $\partial_{x_{i}}u_{i}=0$ and
$\partial_{x_{i}^{\prime}}u_{i}^{\prime}=0$, but no approximations and no
average, gives (Hill 2002a,b)%
\begin{equation}
\partial_{r_{j}}\left[  \left(  u_{i}-u_{i}^{\prime}\right)  \left(
u_{j}-u_{j}^{\prime}\right)  \right]  =\partial_{X_{i}}\left[  \left(
u_{i}+u_{i}^{\prime}\right)  \left(  u_{j}-u_{j}^{\prime}\right)  \right]
/2.\label{exact incomp 2nd order}%
\end{equation}
Apply an ensemble average and the definition (\ref{dij}) in
(\ref{exact incomp 2nd order}). \ Then, to obtain $\partial_{r_{j}%
}\left\langle d_{ij}\right\rangle =0$ from (\ref{exact incomp 2nd order}),
local homogeneity must be the approximation that
\begin{equation}
\partial_{r_{j}}\left\langle d_{ij}\right\rangle \gg\partial_{X_{i}%
}\left\langle \left(  u_{i}+u_{i}^{\prime}\right)  \left(  u_{j}-u_{j}%
^{\prime}\right)  \right\rangle /2.\label{local homo approx}%
\end{equation}
This is a quantifiable approximation because $\left\langle \left(  u_{i}%
+u_{i}^{\prime}\right)  \left(  u_{j}-u_{j}^{\prime}\right)  \right\rangle $
in (\ref{local homo approx}) can be measured at several locations $\mathbf{X}$
such that the derivative on the right-hand side of (\ref{local homo approx})
can be calculated numerically. \ Similarly, local stationarity is the
approximation that the derivatives of statistics with respect to \c{T}\ are
negligible. \ The cases of time and space averages applied to
(\ref{exact incomp 2nd order}) is given in Hill (2002a,b).

\qquad Kolmogorov (1941) introduced a formalism of local homogeneity that uses
the joint probability distribution function (JPDF) of velocity differences.
\ The moment $\left\langle \left(  u_{i}+u_{i}^{\prime}\right)  \left(
u_{j}-u_{j}^{\prime}\right)  \right\rangle $ in (\ref{local homo approx})
cannot be calculated from that JPDF. \ Because of
(\ref{exact incomp 2nd order})--(\ref{local homo approx}), Kolmogorov's
formalism cannot be used to obtain the incompressibility relation
$\partial_{r_{j}}\left\langle d_{ij}\right\rangle =0$, nor is it applicable to
simplifying the structure-function equations deduced from the Navier-Stokes
equation (Hill, 2001, 2002a,b, 2006). \ Kolmogorov's (1941) formalism invokes
a region of vague size for use of the JPDF. \ In contrast,
(\ref{local homo approx}) is truly local because it is a derivative;
experimentalists need only displace $\mathbf{X}$\ sufficiently to use the
3-point numerical derivative formula.

\section{Application of the Navier-Stokes equation for anisotropic turbulence}

\qquad From the Navier-Stokes equations, we obtain an exact equation relating
3rd- and 2nd-order velocity structure functions and other statistics (Hill,
2002a,b, 2006).\ \textquotedblleft Exact\textquotedblright\ means that no
approximations were used; calculus and algebra were used. \ The trace is
performed because it greatly simplifies the term involving the
pressure-gradient difference. \ We obtain%
\begin{equation}
A_{ii}=\partial_{\text{\c{T}}}d_{ii}+\partial_{X_{k}}\digamma_{iik}%
+\partial_{r_{k}}d_{iik}=2\nu\partial_{r_{k}}\partial_{r_{k}}d_{ii}-2\left(
\varepsilon+\varepsilon^{\prime}\right)  +W,\label{trace}%
\end{equation}
where
\begin{equation}
W\equiv-2\partial_{X_{i}}\left[  \left(  p-p^{\prime}\right)  \left(
u_{i}-u_{i}^{\prime}\right)  \right]  +\frac{\nu}{2}\partial_{X_{k}}%
\partial_{X_{k}}d_{ii}-2\nu\partial_{X_{k}}\partial_{X_{k}}\left(
p+p^{\prime}\right)  .\label{Wtrace}%
\end{equation}
No average exists in (\ref{trace})--(\ref{Wtrace}).

\qquad After performing the ensemble average, use of the approximation of
local homogeneity as in \S 3, ie., derivatives with respect to $\mathbf{X}$
are negibible, causes the average of all terms in (\ref{Wtrace}) as well as
the term $\partial_{X_{k}}\left\langle \digamma_{iik}\right\rangle $ to be
negligible because they are all the rate of change of a statistic with respect
to where the measurement is performed, i.e., $\mathbf{X}$. \ Likewise, the
approximation of local stationarity in \S 3 causes the one term $\partial
_{\text{\c{T}}}\left\langle d_{ii}\right\rangle $ to be negligible because it
is the rate of change of $\left\langle d_{ii}\right\rangle $ with respect to
when the measurement is performed, i.e., \c{T}. \ We will return to the
evaluation of those neglected terms in the discussion \S 7.\ \ The result is
the approximate structure-function equation (Hill, 2006).%
\begin{equation}
\left\langle A_{ii}\right\rangle =\partial_{r_{k}}\left\langle d_{iik}%
\right\rangle =2\nu\partial_{r_{k}}\partial_{r_{k}}\left\langle d_{ii}%
\right\rangle -2\left\langle \varepsilon+\varepsilon^{\prime}\right\rangle
.\label{structure function}%
\end{equation}
Note the two equality signs in both (\ref{trace}) and
(\ref{structure function}). \ In (\ref{structure function}) $\partial_{r_{k}%
}\left\langle d_{iik}\right\rangle $\ is the divergence of the vector
$\left\langle d_{iik}\right\rangle $; e.g., in Cartesian coordinates
$\partial_{r_{k}}\left\langle d_{iik}\right\rangle \equiv\partial_{r_{1}%
}\left\langle d_{ii1}\right\rangle +\partial_{r_{2}}\left\langle
d_{ii2}\right\rangle +\partial_{r_{3}}\left\langle d_{ii3}\right\rangle $.
\ The energy dissipation rate in (\ref{trace}) and (\ref{structure function})
is defined by
\begin{equation}
\varepsilon\equiv2\nu s_{ij}s_{ij}.\label{epsilon definition}%
\end{equation}
The calculation of exact averages is described elsewhere.(Hill 2001, 2002a,b,
2006)\ \ Structure functions that contain the two-point pressure difference
present formidable experimental difficulties; it is therefore significant that
the pressure does not appear in (\ref{structure function}) on the basis of
local homogeneity (Hill 2002a,b, 2006). \ Local isotropy was not used to
obtain (\ref{structure function}); therefore, (\ref{structure function}) can
provide methods for measuring the energy dissipation rate in anisotropic turbulence.

\qquad For simplicity of notation, let%
\begin{equation}
\overset{\text{\textbf{\_}}}{\varepsilon}\left(  \mathbf{r}\right)
\equiv\left\langle \varepsilon+\varepsilon^{\prime}\right\rangle
/2.\label{epsilon overbar}%
\end{equation}
Dependence on $\mathbf{r}$\ in (\ref{epsilon overbar}) is because
$\overset{\text{\textbf{\_}}}{\varepsilon}\left(  \mathbf{r}\right)  $ depends
on energy dissipation rates at two points separated by $\mathbf{r}$.
\ Anisotropic turbulence causes dependence of $\overset{\text{\textbf{\_}}%
}{\varepsilon}\left(  \mathbf{r}\right)  $\ on the direction of $\mathbf{r}$;
hence the emphasis on $\mathbf{r}$ in (\ref{epsilon overbar}). \ In
(\ref{epsilon overbar}) the argument list could be $\left(  \mathbf{r,}%
\text{\c{t}}\right)  $, which assumes that the ensemble contains events for
fixed $\mathbf{r}$ and \c{t}. \ However, if an experimenter chooses an
ensemble of events having fixed $\mathbf{r}$\ but various \c{t}, then there is
an implicit average over \c{t} such that \c{t} should be deleted from the
argument list. \ Henceforth, \c{t} is deleted from the argument list.

\qquad For simplicity, consider the case for which $r$\ is much greater than
dissipation scales; for that case we neglect $2\nu\partial_{r_{k}}%
\partial_{r_{k}}\left\langle d_{ij}\right\rangle $ in
(\ref{structure function}), but we later include $2\nu\partial_{r_{k}}%
\partial_{r_{k}}\left\langle d_{ij}\right\rangle $.\ \ Then, because
$A_{ii}\equiv2\left(  a_{i}-a_{i}^{\prime}\right)  \left(  u_{i}-u_{i}%
^{\prime}\right)  $, (\ref{structure function}) gives%
\begin{equation}
\left\langle \left(  a_{i}-a_{i}^{\prime}\right)  \left(  u_{i}-u_{i}^{\prime
}\right)  \right\rangle =-2\overset{\text{\textbf{\_}}}{\varepsilon}\left(
\mathbf{r}\right)  .\label{OMA}%
\end{equation}
For anisotropic turbulence, (\ref{OMA}) makes it clear that $\left\langle
\left(  a_{i}-a_{i}^{\prime}\right)  \left(  u_{i}-u_{i}^{\prime}\right)
\right\rangle $\ depends on the direction of $\mathbf{r}$. \ Mann \textit{et
al.} (1999) obtained a relationship similar to (\ref{OMA}), and they (Mann
\textit{et al.}, 1999, Ott \& Mann, 2000, 2005) used their relationship to
obtain energy dissipation rate from their measurements of acceleration and
velocity. \ For our purpose of mathematical precision, it is necessary to note
the distinctions that (\ref{OMA}) is obtained here and in Hill (2006) without
the assumption by Mann \textit{et al.} (1999) that a certain derivative moment
may be neglected and that both $\varepsilon$ and $\varepsilon^{\prime}$ define
$\overset{\text{\textbf{\_}}}{\varepsilon}\left(  \mathbf{r}\right)  $ in
(\ref{epsilon overbar}), i.e., two space-time points appear here.

\qquad Average (\ref{OMA}) over orientations of $\mathbf{r}$\ to obtain:%
\begin{align}
\frac{1}{4\pi}%
{\displaystyle\iint}
\left\langle \left(  a_{i}-a_{i}^{\prime}\right)  \left(  u_{i}-u_{i}^{\prime
}\right)  \right\rangle d\Omega &  =-2\overset{\text{\textbf{\_}}}%
{\varepsilon}_{\text{orientation}}\left(  r\right) \\
\overset{\text{\textbf{\_}}}{\varepsilon}_{\text{orientation}}\left(
r\right)   &  \equiv\frac{1}{4\pi}%
{\displaystyle\iint}
\overset{\text{\textbf{\_}}}{\varepsilon}d\Omega
\end{align}
which is a function of $r$; $d\Omega$\ is the differential of solid angle, and
the double integral\ is understood to be over $4\pi$ steradians.

\qquad Use of particle tracking data allows calculation of the average of
(\ref{OMA}) over a sphere in $\mathbf{r}$-space to obtain energy dissipation
averaged within the sphere as follows:%
\begin{align}
\frac{3}{4\pi r_{S}^{3}}\underset{\left\vert \mathbf{r}\right\vert \leq r_{S}%
}{%
{\displaystyle\iiint}
}\left\langle \left(  a_{i}-a_{i}^{\prime}\right)  \left(  u_{i}-u_{i}%
^{\prime}\right)  \right\rangle d\mathbf{r} &  \mathbf{=}\frac{3}{r_{S}^{3}%
}\overset{r_{S}}{\underset{0}{%
{\displaystyle\int}
}}\left(  \frac{1}{4\pi}%
{\displaystyle\iint}
\left\langle \left(  a_{i}-a_{i}^{\prime}\right)  \left(  u_{i}-u_{i}^{\prime
}\right)  \right\rangle d\Omega\right)  \text{ }r^{2}\text{ }dr\nonumber\\
&  =-2\overset{\text{\textbf{\_}}}{\varepsilon}_{\text{sphere}}\left(
r_{S}\right)  .
\end{align}%
\begin{equation}
\overset{\text{\textbf{\_}}}{\varepsilon}_{\text{sphere}}\left(  r_{S}\right)
\mathbf{\equiv}\frac{3}{4\pi r_{S}^{3}}\underset{\left\vert \mathbf{r}%
\right\vert \leq r_{S}}{%
{\displaystyle\iiint}
}\overset{\text{\textbf{\_}}}{\varepsilon}d\mathbf{r.}%
\label{epsilon over-bar sphere ave}%
\end{equation}
The sphere has radius $r_{S}$ such that $\overset{\text{\textbf{\_}}%
}{\varepsilon}_{\text{sphere}}$\ depends on $r_{S}$, not on $\mathbf{r}$.
\ The average over orientations of $\mathbf{r}$\ is performed first, resulting
in a function of $r$; the $r$-integration is performed second. \ In
(\ref{epsilon over-bar sphere ave}), the average produces the same result as
that given in the first description of intermittency theory in Obukhov
(1962)\textbf{\ }which was used by Kolmogorov (1962).

\qquad Another method is to use the equality in (\ref{structure function})
that contains $\partial_{r_{k}}\left\langle d_{iik}\right\rangle $. \ To avoid
substituting data into the divergence of the 3rd-order structure function, the
$\mathbf{r}$-sphere\ average is performed and the divergence theorem is used
to express the result as an integral over the surface of the outward normal of
the vector $\left\langle d_{iik}\right\rangle $; then%
\begin{align}
\frac{3}{4\pi r_{S}^{3}}\underset{\left\vert \mathbf{r}\right\vert \leq r_{S}%
}{%
{\displaystyle\iiint}
}\partial_{r_{k}}\left\langle d_{iik}\right\rangle \mathbf{dr} &  =\frac
{3}{4\pi r_{S}^{3}}\underset{\left\vert \mathbf{r}\right\vert =r_{s}}{%
{\displaystyle\iint}
}\frac{r_{k}}{r}\left\langle d_{iik}\right\rangle ds\mathbf{=}\frac{1}{4\pi}%
{\displaystyle\iint}
\left[  \left\langle d_{ii1}\right\rangle \right]  _{\left\vert \mathbf{r}%
\right\vert =r_{s}}d\Omega\nonumber\\
&  =-\frac{4}{3}r_{S}\overset{\text{\textbf{\_}}}{\varepsilon}_{\text{sphere}%
}\left(  r_{S}\right)  .\label{3rd order orientation average}%
\end{align}
The differential of surface area on the sphere is $ds$; $\frac{r_{k}}{r}$ is
the unit vector in the direction of $\mathbf{r}$ such that $\frac{r_{k}}%
{r}\left\langle d_{iik}\right\rangle =\left\langle d_{ii1}\right\rangle $.
\ Specifically, $\left\langle d_{ii1}\right\rangle \equiv\left\langle
d_{111}+d_{221}+d_{331}\right\rangle $. \ The subscript notation $\left[
\left\langle d_{ii1}\right\rangle \right]  _{\left\vert \mathbf{r}\right\vert
=r_{s}}$ means evaluate the quantity within the square brackets at $\left\vert
\mathbf{r}\right\vert =r_{s}$. \ In (\ref{3rd order orientation average}), we
have the orientation average of $\left\langle d_{ii1}\right\rangle $. \ Taylor
\textit{et al.} (2003) use DNS data to demonstrate the efficacy of the
orientation average of the 3rd-order structure function in its relationship to
$\varepsilon$.

\qquad Neglect of the term $\nu\partial_{r_{k}}\partial_{r_{k}}\left\langle
d_{ii}\right\rangle $\ in (\ref{structure function}) is unnecessary. \ The
Laplacian is the divergence of the gradient such that the divergence theorem
gives%
\[
\underset{\left\vert \mathbf{r}\right\vert \leq r_{S}}{%
{\displaystyle\iiint}
}\partial_{r_{k}}\partial_{r_{k}}\left\langle d_{ii}\right\rangle
d\mathbf{r}=\underset{\left\vert \mathbf{r}\right\vert =r_{s}}{%
{\displaystyle\iint}
}\frac{r_{k}}{r}\partial_{r_{k}}\left\langle d_{ii}\right\rangle ds=r_{S}^{2}%
{\displaystyle\iint}
\left[  \partial_{r_{1}}\left\langle d_{ii}\right\rangle \right]  _{\left\vert
\mathbf{r}\right\vert =r_{s}}d\Omega,
\]
where $\partial_{r_{1}}$\ is the gradient in the direction of $\mathbf{r}$.
\ Thus, more generally, for anisotropic turbulence%
\begin{align}
\left\langle \left(  a_{i}-a_{i}^{\prime}\right)  \left(  u_{i}-u_{i}^{\prime
}\right)  \right\rangle -\nu\partial_{r_{k}}\partial_{r_{k}}\left\langle
d_{ii}\right\rangle  &  =-2\overset{\text{\textbf{\_}}}{\varepsilon}\left(
\mathbf{r}\right)  .\label{1 result}\\
\frac{1}{4\pi}%
{\displaystyle\iint}
\left[  \left\langle \left(  a_{i}-a_{i}^{\prime}\right)  \left(  u_{i}%
-u_{i}^{\prime}\right)  \right\rangle -\nu\partial_{r_{k}}\partial_{r_{k}%
}\left\langle d_{ii}\right\rangle \right]  d\Omega &  =-2\overset
{\text{\textbf{\_}}}{\varepsilon}_{\text{orientation}}\left(  r\right)
.\label{2 result}\\
\left\{
\begin{array}
[c]{c}%
\overset{r_{S}}{\underset{0}{%
{\displaystyle\int}
}}\left(  \frac{1}{4\pi}%
{\displaystyle\iint}
\left\langle \left(  a_{i}-a_{i}^{\prime}\right)  \left(  u_{i}-u_{i}^{\prime
}\right)  \right\rangle d\Omega\right)  r^{2}dr\\
-\frac{\nu}{4\pi}r_{S}^{2}%
{\displaystyle\iint}
\left[  \partial_{r_{1}}\left\langle d_{ii}\right\rangle \right]  _{\left\vert
\mathbf{r}\right\vert =r_{s}}d\Omega
\end{array}
\right\}   &  =-\frac{2}{3}r_{S}^{3}\overset{\text{\textbf{\_}}}{\varepsilon
}_{\text{sphere}}\left(  r_{S}\right)  .\label{3 result}\\
\frac{1}{4\pi}%
{\displaystyle\iint}
\left[  \left\langle d_{ii1}\right\rangle -2\nu\partial_{r_{1}}\left\langle
d_{ii}\right\rangle \right]  _{\left\vert \mathbf{r}\right\vert =r_{s}}d\Omega
&  =-\frac{4}{3}r_{S}\overset{\text{\textbf{\_}}}{\varepsilon}_{\text{sphere}%
}\left(  r_{S}\right)  .\label{4a result}%
\end{align}
For stationary, homogeneous turbulence, (\ref{1 result})--(\ref{4a result})
are valid for all\ $r$\ and all Reynolds numbers. \ We began with the
hydrodynamic definition of energy dissipation rate, $\varepsilon$, in
(\ref{epsilon definition}); (\ref{epsilon overbar}) is only simplified
notation. \ We obtain in (\ref{1 result})--(\ref{4a result})\ three different
averages of $\varepsilon$. \ The three values differ only because the
averaging operations differ. \ There is no distinction as to which value to
prefer. \ Any choice of averaging operation is left to the judgment of the experimenter.

\section{The limit of local isotropy}

\qquad We can determine whether or not the above formulas give the
corresponding classic results for locally isotropic turbulence. \ All of the
energy dissipation rates in (\ref{1 result})--(\ref{4a result}) become the
same value for local isotropy, so here we denote them all by $\left\langle
\varepsilon\right\rangle $. \ For local isotropy, both (\ref{1 result}) and
(\ref{2 result}) become%
\begin{equation}
\left\langle \left(  a_{i}-a_{i}^{\prime}\right)  \left(  u_{i}-u_{i}^{\prime
}\right)  \right\rangle -\nu r^{-2}\partial_{r}\left(  r^{2}\partial
_{r}\right)  \left\langle d_{ii}\right\rangle =-2\varepsilon,\label{isotropy1}%
\end{equation}
where $\partial_{r_{k}}\partial_{r_{k}}=r^{-2}\partial_{r}\left(
r^{2}\partial_{r}\right)  $ was used, and (\ref{3 result})--(\ref{4a result})
become
\begin{align}
\overset{r}{\underset{0}{%
{\displaystyle\int}
}}\left\langle \left(  a_{i}-a_{i}^{\prime}\right)  \left(  u_{i}%
-u_{i}^{\prime}\right)  \right\rangle r^{2}dr-\nu r^{2}\partial_{r}%
\left\langle d_{ii}\right\rangle  &  =-\frac{2}{3}r^{3}\varepsilon
.\label{isotropy 2}\\
\left\langle d_{ii1}\right\rangle -2\nu\partial_{r}\left\langle d_{ii}%
\right\rangle  &  =-\frac{4}{3}r\varepsilon.\label{isotropy3}%
\end{align}
The inertial range formulas obtained from (\ref{isotropy1})--(\ref{isotropy3})
are the well-known formulas
\[
\left\langle \left(  a_{i}-a_{i}^{\prime}\right)  \left(  u_{i}-u_{i}^{\prime
}\right)  \right\rangle =-2\left\langle \varepsilon\right\rangle \text{ \ and
\ }\left\langle d_{ii1}\right\rangle =-\frac{4}{3}r\left\langle \varepsilon
\right\rangle
\]
The viscous-range formulas from (\ref{isotropy1})--(\ref{isotropy3}) are $\nu
r^{-2}\partial_{r}\left(  r^{2}\partial_{r}\right)  \left\langle
d_{ii}\right\rangle =2\left\langle \varepsilon\right\rangle $ and $\nu
\partial_{r}\left\langle d_{ii}\right\rangle =\frac{2}{3}r\left\langle
\varepsilon\right\rangle $. \ In the viscous range, Taylor series expansion,
local isotropy, and incompressibility give%
\begin{equation}
\left\langle d_{ii}\right\rangle =\left\langle \left(  \frac{\partial u_{1}%
}{\partial x_{1}}\right)  ^{2}+2\left(  \frac{\partial u_{2}}{\partial x_{1}%
}\right)  ^{2}\right\rangle r^{2}=5\left\langle \left(  \frac{\partial u_{1}%
}{\partial x_{1}}\right)  ^{2}\right\rangle r^{2}=\frac{5}{2}\left\langle
\left(  \frac{\partial u_{2}}{\partial x_{1}}\right)  ^{2}\right\rangle r^{2}.
\end{equation}
Note that $r^{-2}\partial_{r}\left(  r^{2}\partial_{r}\right)  r^{2}=6$.
\ Then, from (\ref{isotropy1})--(\ref{isotropy3}), both $\nu r^{-2}%
\partial_{r}\left(  r^{2}\partial_{r}\right)  \left\langle d_{ii}\right\rangle
=2\left\langle \varepsilon\right\rangle $ and $\nu\partial_{r}\left\langle
d_{ii}\right\rangle =\frac{2}{3}r\left\langle \varepsilon\right\rangle $ give
the classic formulas%
\begin{equation}
\left\langle \left(  \frac{\partial u_{1}}{\partial x_{1}}\right)
^{2}\right\rangle =\frac{1}{15\nu}\left\langle \varepsilon\right\rangle
\text{, \ \ \ and \ \ \ }\left\langle \left(  \frac{\partial u_{2}}{\partial
x_{1}}\right)  ^{2}\right\rangle =\frac{2}{15\nu}\left\langle \varepsilon
\right\rangle .\label{isotropic derivative formula}%
\end{equation}

\section{Discussion}

\qquad The mathematics is precise. \ Experimenters must be careful to follow
with precision when using the equations. \ In particular, precise evaluation
of $\varepsilon$ requires evaluation of not only (\ref{1 result}%
)--(\ref{4a result}), but also the terms in (\ref{trace}) that describe
inhomogeneity, i.e., $\partial_{X_{k}}\digamma_{iik}$ and $W$, as well as the
term that describes nonstationarity, i.e., $\partial_{\text{\c{T}}}d_{ii}$.
\ Those terms must be operated upon with the same averages that appear in
(\ref{1 result})--(\ref{4a result}). \ Danaila \textit{et al}. (2002, 2004,
and references therein) give approximate evaluations of some of those terms
for several flows. \ The pressure that appears in $W$ in (\ref{Wtrace})
presents a future challenge, but techniques of pressure measurement combined
with hot-wire velocity measurement have advanced (Tsuji \textit{et al}., 2007).

\qquad The same velocity and acceleration that appear in the Navier-Stokes
equation also appear above. \ That is, the velocity fluctuation and
acceleration fluctuation do not appear above. \ If the experimenter performs a
Reynolds decomposition, e.g. $u=\left\langle u\right\rangle +\widetilde{u}$
where $\widetilde{u}$ is the fluctuation of velocity, then many more terms
must appear. \ Those terms must be evaluated quantitatively to obtain
quantitative energy dissipation rate. \ The energy dissipation rate may be
expressed as being caused by the average flow and the fluctuation flow and
terms descriptive of the interaction of the two. \ The subject of Reynolds
decomposition for the structure-function equations, and the many resultant
terms that must be quantified or neglected is discussed in Hill (2002b). \ If
those many terms are neglected, the resultant approximate equations for
structure functions of fluctuations are described fully in Hill (2002b).

\qquad An important advantage of using $\left\langle A_{ii}\right\rangle
$\ over $\partial_{r_{k}}\left\langle d_{iik}\right\rangle $ in
(\ref{structure function}) is the necessity to neglect $\partial
_{\text{\c{T}}}\left\langle d_{ii}\right\rangle +\partial_{X_{k}}\left\langle
\digamma_{iik}\right\rangle $\ in (\ref{trace}) if $\partial_{r_{k}%
}\left\langle d_{iik}\right\rangle $ is to be used. \ For the case of a
nonzero mean velocity as well as for random sweeping by the large-scale flow,
the term $\partial_{X_{k}}\left\langle \digamma_{iik}\right\rangle $\ can be a
significant effect [see \S 7 of Hill (2006) and experimental evaluation in
Danaila \textit{et al}. (2002, 2004, and references therein)].

\qquad The definition of local homogeneity used above holds at the center of
symmetry of a flow. \ For example, in the center of the cylinder containing
the flow between rotating blades, the rate of change with respect to where the
measurement is performed (i.e., $\mathbf{X}$) is zero because any statistic
evaluated both above and below the center of symmetry has the same value at
both points. The same is true for any statistic evaluated at two points
equally spaced in any direction from the center of symmetry.

\qquad For use of Taylor's hypothesis of frozen flow, the correction for
fluctuating convection velocity is given in Hill (1996) for any statistic on
the basis of local isotropy; that makes that correction inapplicable to
anisotropic turbulence. \ However, the correction is qualitatively useful.
\ An advantage of the 3rd-order structure function $\left\langle
d_{ijk}\right\rangle $ is that the correction to Taylor's hypothesis caused by
fluctuating velocity vanishes in the inertial range of $\left\langle
d_{ijk}\right\rangle $ (Hill, 1996); the same can be shown to be true for
$\left\langle A_{ij}\right\rangle $.

\qquad In (\ref{1 result})--(\ref{2 result}) the calculation of the
three-dimensional Laplacian of data might pose problems. \ Danaila \textit{et
al}. (2002, 2004, and references therein) have approximated the Laplacian by
only its $r$-derivatives on the basis that local isotropy should be adequate
at the dissipation and viscous-range scales\ where $\nu\partial_{r_{k}%
}\partial_{r_{k}}\left\langle d_{ii}\right\rangle $ is an important term
within (\ref{1 result})--(\ref{2 result}). \ For axisymmetric turbulence one
may express the Laplacian in terms of only two independent variables using the
cylindrical coordinate system used by Lindborg (1995) or the coordinates used
by Batchelor (1946). \ The axisymmetric analogues of
(\ref{isotropic derivative formula}) are given by George and Hussein (1991).

\qquad The results (\ref{1 result})--(\ref{4a result}) relate the energy
dissipation rates on the right-hand sides to quantities that are measurable on
the left-hand sides. \ The results are valid for anisotropic turbulence. \ The
left-hand sides of (\ref{1 result})--(\ref{4a result}) are measurable with
particle tracking technology (Voth \textit{et al.} 2002, La Porta \textit{et
al.} 2001, Mann \textit{et al.} 1999, Ott \& Mann 2000, 2005, Luthi \textit{et
al.}, 2005, Berg \textit{et al.} 2006, Ouellette \textit{et al.} 2006). \ With
the exception of the acceleration-velocity structure function, the left-hand
sides of (\ref{1 result})--(\ref{4a result}) are also measurable by multi-wire
probes (Tsinober \textit{et al.} 1992, Kolmyansky \textit{et al.} 2001,
Gulitski \textit{et al.} 2007). \ We expect that (\ref{1 result}%
)--(\ref{4a result}) will be of significant use to experimenters.

\ \ The author thanks A. Pumir, B. Luthi, H. Xu \& E.\ Bodenschatz for
valuable comments.

\ \ \ \ \ \ \ \ \ \ \ \ \ \ \ \ \ \ \ \ \ \ \ \ \ \ \ \ \ \ \ \ \ \ \ \ REFERENCES

\noindent\textsc{Batchelor, G. K.} 1946 The theory of axisymmetric turbulence.

\qquad\textit{Proc. R. Soc. Lond.} A \textbf{186}, 480--502.

\noindent\textsc{Berg, J., Luthi, B., Mann J. \& Ott, S.} 2006\ Backwards and
forwards relative

\qquad dispersion in turbulent flow: \ An experimental investigation.
\textit{Phys. Rev. E} \textbf{74}(1),

\qquad016304.

\noindent\textsc{Danaila, L., Anselmet, F. \& Antonia, R. A.} 2002\ An
overview of the effect of

\qquad large-scale inhomogeneities on small-scale turbulence. \textit{Phys.
Fluids }\textbf{14}, 2475--2484.

\noindent\textsc{Danaila, L., Antonia, R. A. \& Burattini, P.} 2004\ Progress
in studying small-scale

\qquad turbulence using `exact' two-point equations. \textit{New J. Phys.
}\textbf{6}, 128.

\noindent\textsc{Duchon, J. \& Robert, R.} 2000\ Inertial energy dissipation
for weak solutions

\qquad of incompressible Euler and Navier-Stokes equations.
\textit{Nonlinearity} \textbf{13}, 229--255.

\noindent\textsc{Eyink, G.} 2003\ Local 4/5-law and energy dissipation anomaly
in turbulence.

\qquad\textit{Nonlinearity} \textbf{16}, 137--145.

\noindent\textsc{George, W. K., \& Hussein, H. J.,} 1991\ Locally axisymmetric turbulence.

\qquad\textit{J.~Fluid Mech.} \textbf{233}, 1--23.

\noindent\textsc{Gulitski G., Kolmyansky, M., Kinzelbach, W., Luthi,
B.,\ Tsinober, A.,}

\qquad\textsc{\& Yorish, S.}\ 2007 Velocity and temperature derivatives in
high-Reynolds-number turbulent flows

\qquad in the atmospheric surface layer. \ Part 2, Accelerations and related matters

\qquad\textit{J.~Fluid Mech.} \textbf{589}, 83--102.

\noindent\textsc{Hill, R. J.} 1996 Corrections to Taylor's frozen turbulence approximation.

\qquad\textit{Atmos. Res. }\textbf{40}, 153--175.

\noindent\textsc{Hill, R. J.} 2001 Equations relating structure functions of
all orders. \textit{J.~Fluid Mech.}

\qquad\textbf{434}, 379--388.

\noindent\textsc{Hill, R. J.} 2002a Exact second-order structure-function
relationships. \textit{J.~Fluid Mech.}

\qquad\textbf{468}, 317--326.

\noindent\textsc{Hill, R. J.} 2002b The approach of turbulence to the locally
homogeneous asymptote as

\qquad studied using exact structure-function equations. (xxx.lanl.gov/physics/0206034).

\noindent\textsc{Hill, R. J.} 2006 Opportunities for use of exact statistical equations.

\qquad\textit{J.~Turbulence.} \textbf{7}, No. 43, pp.13.

\noindent\textsc{Kolmogorov, A. N.} 1941 The local structure of turbulence in
incompressible viscous

\qquad fluid for very large Reynolds numbers. \textit{Doklady Akademia Nauk
SSSR} \textbf{30}, 301--305

\noindent\textsc{Kolmogorov, A. N.} 1962\ A refinement of previous hypotheses
concerning the

\qquad local structure of turbulence in a viscous incompressible fluid at high Reynolds

\qquad number. \textit{J.~Fluid Mech.} \textbf{13}, 82--85.

\noindent\textsc{Kolmyansky, M., Tsinober, A. \& Yorish, S.}\ 2001\ Velocity
derivatives in the

\qquad atmospheric surface layer at $Re_{\lambda}=10^{4}$. \textit{Phys.
Fluids} \textbf{13}, 311--314.

\noindent\textsc{La Porta, A., Voth, G. A., Crawford, A. M., Alexander, A. \&
Bodenschatz,}

\qquad\textsc{E.}\ 2001\ Fluid particle accelerations in fully developed
turbulence. \textit{Nature}

\qquad\textbf{409}, 1017--1019.

\noindent\textsc{Lindborg, E.} 1995 Kinematics of homogeneous axisymmetric turbulence.

\qquad\textit{J.~Fluid Mech.} \textbf{302}, 179--201.

\noindent\textsc{Luthi, B., Tsinober, A., \& Kinzelbach, W.,} 2005\ Lagrangian
measurement of

\qquad vorticity dynamics in turbulent flows.\ \textit{J.~Fluid Mech.}
\textbf{528}, 87--118.

\noindent\textsc{Mann, J., Ott, S. \& Anderson, J. S.} 1999\ Experimental
study of relative turbulent

\qquad diffusion. \textit{Technical Report Riso-R-1036(EN)}

\qquad(Roskilde, Denmark: Riso National Laboratory).

\noindent\textsc{Nie, Q. \& Tanveer, S.} 1999\ A note on third-order structure
functions in turbulence.

\qquad Proc.\textit{\ R. Soc. Lond. A,} \textbf{455}, 1615--1635.

\noindent\textsc{Obukhov, A. M.} 1962 Some specific features of atmospheric
turbulence. \textit{J.~Fluid Mech.}

\qquad\textbf{13}, 77--81.

\noindent\textsc{Ott, S. \& Mann, J.} 2000\ An experimental investigation of
relative dispersion of particle

\qquad pairs in three-dimensional turbulent flow. \textit{J.~Fluid Mech.}
\textbf{422}, 201--223.

\noindent\textsc{Ott, S. \& Mann, J.} 2005\ An experimental test of Corrsin's
conjecture and some related

\qquad ideas. \textit{New J.~Phys.}\textbf{\ 7}, 142.

\noindent\textsc{Ouellette, N. T., Xu, H., Bourgoin, M., \& Bodenschatz, E.}\ 2006\ An

\qquad experimental study of turbulent relative dispersion models. \textit{New
J. Phys. }\textbf{8}, 109.

\noindent\textsc{Taylor, M. A., Kurien, S., Eyink, G. L.} 2003\ Recovering
isotropic statistics in

\qquad turbulence simulations: \ The Kolmogorov 4/5th law.\ \textit{Phys. Rev.
E} \textbf{68}(2), 026310.

\noindent\textsc{Tsinober, A., Kit, E. \& Dracos, T.} 1992\ Experimental
investigation of the field of

\qquad velocity gradients in turbulent flows. \textit{J.~Fluid Mech.}
\textbf{242}, 169--192.

\noindent\textsc{Tsuji, Y., Fransson, J. H. M., Alfredsson, P. H., \&
Johansson, A. V.} 2007

\qquad Pressure statistics and their scaling in high-Reynolds-number boundary layers.

\qquad\ \textit{J.~Fluid Mech.} \textbf{585}, 1--40.

\noindent\textsc{Voth, G. A., LaPorta, A., Crawford, A. M., Alexander, J. \&
Bodenschatz,}

\qquad\textsc{E.} 2002\ Measurement of particle accelerations in fully
developed turbulence.

\qquad\textit{J.~Fluid Mech.} \textbf{469}, 121--160.

\end{document}